\newcommand\beq{\begin{equation}}
\newcommand\eeq{\end{equation}}

\documentclass[12pt,preprint]{aastex}

\shorttitle{Cosmic Ray and WMAP data correlation} \shortauthors{}

\begin{document}

\title{Correlation Analysis between Tibet AS-$\gamma$ TeV Cosmic Ray and WMAP Nine-year Data}

\author{Qian-Qing Yin\altaffilmark{1,2} and Shuang-Nan Zhang\altaffilmark{1,3}}

\email{zhangsn@ihep.ac.cn}

\altaffiltext{1}{Key Laboratory of Particle Astrophysics, Institute of High Energy Physics,
	Beijing 100049, China}

\altaffiltext{2}{University of Chinese Academy of Sciences,
	Beijing 100049, China}

\altaffiltext{3}{National Astronomical Observatories, Chinese Academy of Sciences,
	Beijing 100012, China}

\begin{abstract}

The Wilkinson Microwave Anisotropy Probe (WMAP) team subtracted template-based foreground models to produce foreground-reduced maps, and masked point sources and uncertain sky regions directly; however, whether foreground residuals exist in the WMAP foreground-reduced maps is still an open question. Here, we use Pearson correlation coefficient (PCC) analysis with AS-$\gamma$ TeV cosmic ray (CR) data to probe possible foreground residuals in the WMAP nine-year data. The correlation results between the CR and foreground-contained maps (WMAP foreground-unreduced maps, WMAP template-based and MEM foreground models) suggest that: (1) CRs can trace foregrounds in the WMAP data; (2) at least some TeV CRs originate from the Milky Way; (3) foregrounds may be related to the existence of CR anisotropy (loss-cone and tail-in structures); (4) there exist differences among different types of foregrounds in the declination range of $< 15^{\circ}$. Then, we generate 10,000 mock CMB sky maps to describe the cosmic variance, which is used to measure the effect of the fluctuations of all possible CMB maps to the correlations between CR and CMB maps. Finally, we do correlation analysis between the CR and WMAP foreground-reduced maps, and find that: (1) there are significant anticorrelations; and (2) the WMAP foreground-reduced maps are credible. However, the significant anticorrelations may be accidental, and the higher signal-to-noise ratio (SNR) Planck SMICA map cannot reject the hypothesis of accidental correlations. We therefore can only conclude that the foreground residuals exist with $\sim$ 95\% probability.

\end{abstract}


\section{INTRODUCTION}

The Wilkinson Microwave Anisotropy Probe (hereafter WMAP) is designed to obtain a full sky map of the temperature anisotropy of the cosmic microwave background radiation (CMB) and has provided us with quite useful information in cosmology research \citep{hin13}. The WMAP frequency bands are 22, 33, 41, 61 and 94 GHz, named as K, Ka, Q, V, and W bands, and the corresponding angular resolutions are 0.88, 0.66, 0.51, 0.35, and 0.22 deg, respectively. The WMAP team combines the CMB-dominated bands (Q, V and W bands) to obtain a signal-to-noise ratio (SNR) improved map; however, foregrounds as well as point sources also contribute to the map at these selected bands. A linear combination of foreground templates is fit to the WMAP sky map at each frequency band, and the fit is subtracted to produce the reduced foreground map; the foreground templates are the WMAP K-Ka temperature difference map \citep{hin07}, the H$\alpha$ map \citep{fin03} and the FDS 94 GHz dust emission map \citep{fin99}. Moreover, some techniques, including internal linear combination (ILC), Maximum Entropy Method (MEM), Markov Chain Monte Carlo (MCMC) and $\chi^{2}$, are also implemented to characterize the nature of the foregrounds (synchrotron, free-free, spinning dust and thermal dust). Then averaged foreground models (MEM, MCMCg, and $\chi^{2}$ Model 9) show that the total WMAP foreground emissions dominate at lower frequencies and decrease with increasing frequency till V and W bands (see Figure 22 in \citealt{ben13}). Point sources and sky regions with uncertain foregrounds (generally close to the Galactic plane) can be masked directly from these WMAP maps \citep{ben13}.

Ever since the release of the WMAP first-year data, the foreground residuals in the WMAP foreground-reduced data have received much attention. Joint multiwavelength correlation analysis is an important method for this purpose. Many kinds of data have been used, such as the 10 and 15 GHz Tenerife data \citep{deo99,deo04}, the thermal dust and spinning dust templates \citep{fin04}, the interstellar neutral hydrogen data \citep{ver07}, the Sloan Digital Sky Survey data \citep{pad05,hir08}, the diffuse gamma-ray intensity maps \citep{wib05,liu06}, etc.

Some basic components of the interstellar medium (ISM), which contain Galactic magnetic field (GMF), Galactic cosmic ray (GCR) and the ordinary matter (such as gas and dust), have comparable pressures and are coupled together by electromagnetic forces. Both the dynamics of the ordinary matter and its spatial distribution at all scales are influenced by GCRs and GMFs; the GMFs and the propagation of GCRs provide efficient support against the self-gravity of ordinary matter. Conversely, the inertia of the ordinary matter confines GMFs and, hence, GCRs, while its turbulent motions are responsible for the amplification of GMFs and the acceleration of GCRs \citep{fer01}. The WMAP foreground emissions are closely related to these components of the ISM: the rapid spiraling motions of GCR electrons about magnetic field lines generate the synchrotron emissions \citep{gin65}; the free-free emissions arise from ionized gas; and the spinning dust and thermal dust emissions are obviously from interstellar dust. \cite{wib05} found that there are foreground residuals in the WMAP first-year data, which they interpreted as related to cosmic ray (CR) intensity. \cite{liu06} also found that there are residuals in the WMAP first-year foreground cleaned data. They correlated the WMAP first-year foreground-reduced data with the diffuse gamma-ray intensity maps from the Energetic Gamma-ray Experiment Telescope (EGRET), and found that there are significant correlations at scales around $15^{\circ}$ in Q and W bands. They concluded that the correlations are caused by the foreground residuals in the WMAP data, which may be induced by cosmic rays because at least some diffuse gamma-ray emissions are believed to be produced by the interactions of cosmic rays with gas and photon fields in the Milky Way. Hence, direct correlation analysis between the CR data and the WMAP data may be helpful to understand the nature of foreground residuals in the WMAP data.

The WMAP team has released data several times; the latest and final products are the WMAP nine-year data. The TeV CR intensity maps can be used to do correlation analysis with the WMAP nine-year data. Several TeV CR experiments have been conducted in recent decades, including Super-Kamiokande \citep{gui07}, MILAGRO \citep{abd09}, ARGO-YBJ \citep{zha10}, EAS-TOP \citep{agl09}, IceCube \citep{abb11}, etc. These experiments have published scientific results, including CR anisotropy. The CR data we use here are from the large data sample of Tibet Air Shower Arrays (AS-$\gamma$) experiment at 4, 6, 12, 50 and 300 TeV, respectively; the arrays have been operated at Yangbajing ($90.522^{\circ}$ E, $30.102^{\circ}$ N, 4,300 m above sea level) in Tibet, China since 1990. The AS-$\gamma$ experiment has an large effective area (up to 36,900 $m^{2}$) and provides us enough GCR events (here $\sim$ 37 billions); here the experiment cannot distinguish gamma rays from the charged CR background \citep{ame06}, but cosmic rays dominate over gamma rays anyway in the Tibet air shower data. In our paper the original CR relative intensity maps before smoothing are the same as in Figure 3 of \cite{ame06}, except that unity is subtracted from the intensity. There are two known structures of anisotropy on the CR map, i.e., the loss-cone structure and the tail-in structure \citep{nag76,nag98}. The loss-cone structure is the region with lower CR intensity and looks like a circle on the CR map, while the CR intensity has enhancement in the tail-in region. \cite{ame06} concluded that these anisotropic structures are not related to the heliospheric magnetic field but modulated by GMF, since they exist at energies of a few tens of TeV where the gyroradius is $\gg$ 1 AU. \cite{gia12} showed that such structures arise from some structures in the turbulent GMF within the CR scattering length (i.e., a few tens of parsecs from the Earth). \cite{dru08} suggested that such structures could result from more distant sources if the magnetic mirrors allow for relatively scatter-free CR propagation from the source to the solar system. Hence, the CR anisotropy most likely originates from some local structures in close proximity to the solar system. In Figure 3 of \cite{ame06}, there is a newly discovered structure, the Cygnus region. All these structures change with the CR energy bands, and fade at higher CR energy bands \citep{ame06}; for example, the CR maps at 300 TeV do not show any significant anisotropy. These anisotropic structures are important probes of foreground residuals in the WMAP data.

Planck is the third generation of space-based CMB observatory operated by the European Space Agency, and has released its products in 2013. Compared with the WMAP, the Planck has wider frequency range (25-1000 GHz) and smaller angular resolution (5-30 arcminutes), and provides higher SNR temperature sky map \citep{pla13}. The correlation analysis between the Planck data and the CR data may help to determine whether foreground residuals in the WMAP data exist.

The paper is organized as follows: in Section \ref{sec2} we describe the pre-processing of sky maps and the correlation analysis method. In Section \ref{sec3} we prove that the CR data can be used to probe the WMAP foregrounds. In Section \ref{sec4} we analyse the correlations between the CR data and the WMAP foreground-reduced data, and discuss the probabilities of foreground residuals' existence. Finally, we summarize and make discussion in section \ref{sec5}.

\section{CORRELATION ANALYSIS METHOD\label{sec2}}

The correlation analysis between two maps needs a method to calculate correlation coefficients, and the Pearson correlation coefficient (PCC) method is commonly used. However, the sky maps detected by different instruments may be different in data formats and image qualities; for example, the WMAP maps at different frequency bands correspond to different angular resolutions. Before correlation analysis, we need to make maps with the same format and quality.

\subsection{Pre-Processing of Sky Maps\label{sec21}}

We process all maps by three main steps, including coordinate transformation, map masking and smoothing. The AS-$\gamma$ CR sky maps sample data over square grids of size with $2^{\circ}$ in the declination range of about $-10^{\circ}$ to $70^{\circ}$, while other all-sky maps (WMAP, Planck and foregrounds) are stored with much smaller HEALPix grids. Hence, a practical option is that all maps should be transformed into the AS-$\gamma$ style map. In addition, the WMAP team uses masking templates to mask point sources and sky regions with uncertain foregrounds; here the WMAP Extended Temperature Data Analysis Mask (i.e., the KQ75y9 in \citealt{ben13}) is chosen to cut as many uncertain regions as possible. After masking all maps, we find that the tail-in regions are partly masked except the declination range of less than $15^{\circ}$ and the Cygnus regions are masked totally in the CR maps. Finally, all maps are smoothed to the same angular resolution (= $5^{\circ}$) by Guassian smoothing.
The FWHM of the Gaussian smoothing is calculated by using
\beq {\rm FWHM}=\sqrt{{\rm FWHM}_0^2-{\rm FWHM}_i^2}, \label{fwhm}
\eeq where FWHM$_0$ is the angular resolution to be smoothed to, FWHM$_i$ is the angular resolution of map $i$.

\subsection{Pearson Correlation Coefficient Method\label{sec22}}

The PCC method is used to do correlation analysis. It is defined as
\beq
r_{xy}^{\rm \uppercase\expandafter{\romannumeral1}}= \frac{n\sum_i{x_iy_i}
-\sum_i{x_i}\sum_i{y_i}}{\sqrt{n\sum_i{x^2_i}
-(\sum_i{x_i})^2}\sqrt{n\sum_i{y^2_i}-(\sum_i{y_i})^2}}, \label{rxy}
\eeq
where $x$ and $y$ are the two data samples (e.g., the CR data and the WMAP data), $n$ is the total number of elements in each sample,
and $r_{xy}^{\rm \uppercase\expandafter{\romannumeral1}}$ is the PCC of $x$ (CR) and $y$ (WMAP). Masked pixels by KQ75y9 are not used in this correlation analysis. The values of PCCs range from -1 to 1, and there is no correlation when $r_{xy}^{\rm \uppercase\expandafter{\romannumeral1}}$ = 0.

The errors related to $r_{xy}^{\rm \uppercase\expandafter{\romannumeral1}}$ are obtained by using 10,000 Monte Carlo simulations that considering the errors of $x$ and $y$. $r_{xy}^{\rm \uppercase\expandafter{\romannumeral1}}$ is transformed to $z_{xy}^{\rm \uppercase\expandafter{\romannumeral1}}$ by the Fisher transformation equation:
\beq
z_{xy}^{\rm \uppercase\expandafter{\romannumeral1}}=0.5{\rm ln}\frac{1+r_{xy}^{\rm \uppercase\expandafter{\romannumeral1}}}{1-r_{xy}^{\rm \uppercase\expandafter{\romannumeral1}}}.
\label{zxy}
\eeq
Then $z_{xy}^{\rm \uppercase\expandafter{\romannumeral1}}$ approximately follows a normal distribution with the standard deviation $\sigma_{z_{xy}}^{\rm \uppercase\expandafter{\romannumeral1}}$ (i.e., the error of $z_{xy}^{\rm \uppercase\expandafter{\romannumeral1}}$). Hence, the error of the correlation coefficient $\sigma_{r_{xy}}^{\rm \uppercase\expandafter{\romannumeral1}}$ can be calculated with the inverse Fisher transformation.

By the above pre-processing and PCC method, we can obtain the curves of $r_{xy}^{\rm \uppercase\expandafter{\romannumeral1}}$ (between two sky maps) with declinations; here a correlation coefficient is determined independently for each declination bin ($\sim$ $2^{\circ}$) because absolute calibration
of the Tibet air shower detector efficiency in declination is not available \citep{ame06}. These curves actually measure the correlation of the two sky maps in the spatial distribution, because the positive or negative values of $r_{xy}^{\rm \uppercase\expandafter{\romannumeral1}}$ correspond to correlations or anticorrelations in the spatial distribution that the two maps have; here the correlation or anticorrelation increases with increasing absolute value of $r_{xy}^{\rm \uppercase\expandafter{\romannumeral1}}$, and the two maps have no correlation when $r_{xy}^{\rm \uppercase\expandafter{\romannumeral1}}$ = 0. Besides, if the sky maps for correlation analysis are isotropic, then the correlation coefficient $r_{xy}^{\rm \uppercase\expandafter{\romannumeral1}}$ = 0; therefore, the changes of spatial distributions (e.g., anisotropic structures in the CR maps) are very important. There are two explanations for the observed correlation: two maps are all effected by the same physical processes, so the correlation is causal; or, if the two maps have no any physical relationship, the correlation is accidental. It is important to distinguish between these two possibilities.

\section{CORRELATION BETWEEN CR AND FOREGROUNDS\label{sec3}}

In this section, we study if the CR data can be used as the probe of foreground residuals. We therefore need to do correlation analysis between the CR maps and some sky maps with foregrounds, i.e., the WMAP foreground-unreduced maps, the WMAP template-based foreground models and the MEM foreground models.

\subsection{Correlation betweeen CR and WMAP Data\label{sec31}}

\begin{figure}[H]
\centering
\includegraphics[scale=0.8]{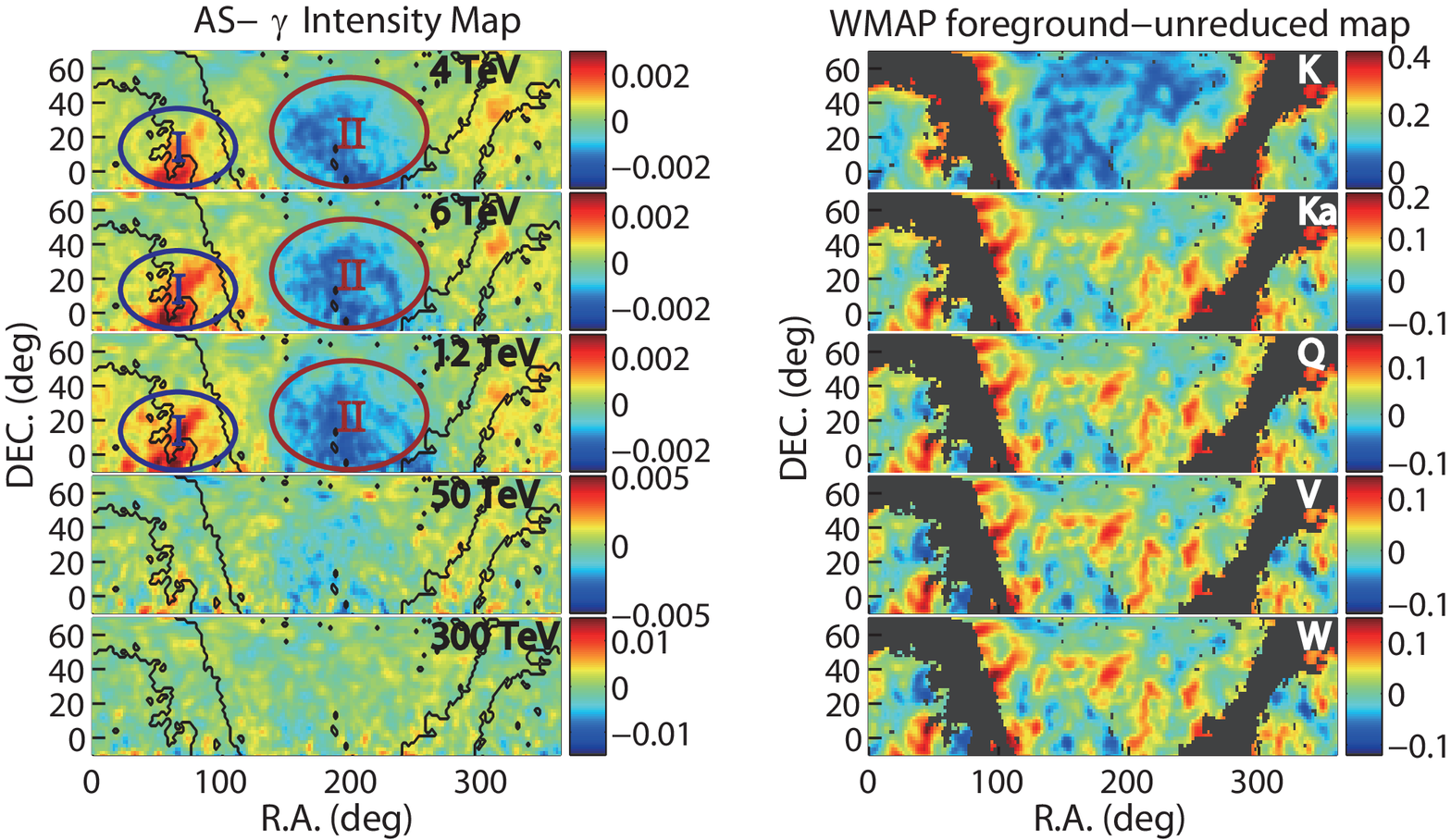}
\caption{Smoothed AS-$\gamma$ cosmic ray (CR) intensity maps and WMAP nine-year
foreground-unreduced temperature fluctuation maps in equatorial
coordinates, sampled over a square grid of size width $2^{\circ}$; all maps are smoothed to $5^{\circ}$.
The circled regions labeled by I and II in the left top three panels are the tail-in regions and the loss-cone regions, respectively. The right five panels are smoothed WMAP
nine-year foreground-unreduced temperature fluctuation maps, which
are masked by the KQ75y9 and the point source catalog masks, in units of
mK; these masks are shown in dark grey in the WMAP foreground-unreduced maps and contours of their boundaries are added to the CR maps.
During the transformation from Galactic coordinates to equatorial
coordinates, a $2^{\circ}\times 2^{\circ}$ grid will be set to zero and not be used in the correlation analysis if
more than 50\% of HEALPix pixels in the grid are masked. \label{Fig1}}
\end{figure}

\begin{figure}[H]
\centering
\includegraphics[scale=0.8]{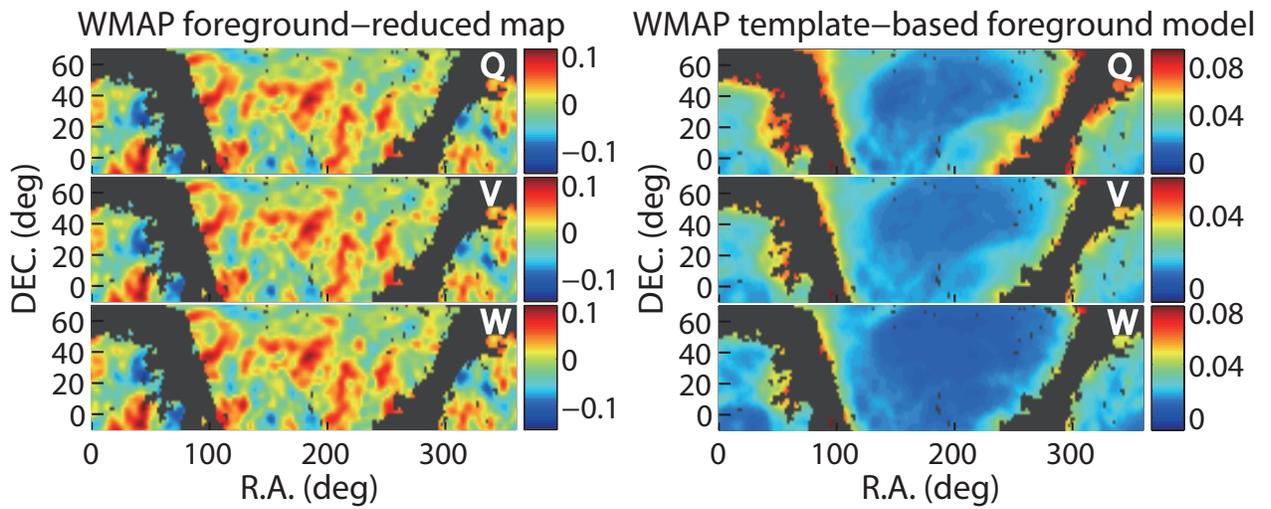}
\caption{Smoothed WMAP nine-year foreground-reduced temperature
fluctuation maps and WMAP template-based foreground models at Q, V, W bands, in
units of mK. The smoothing and mask procedures are the same as Figure \ref{Fig1}. \label{Fig2}}
\end{figure}

\begin{figure}[H]
\centering
\includegraphics[scale=0.8]{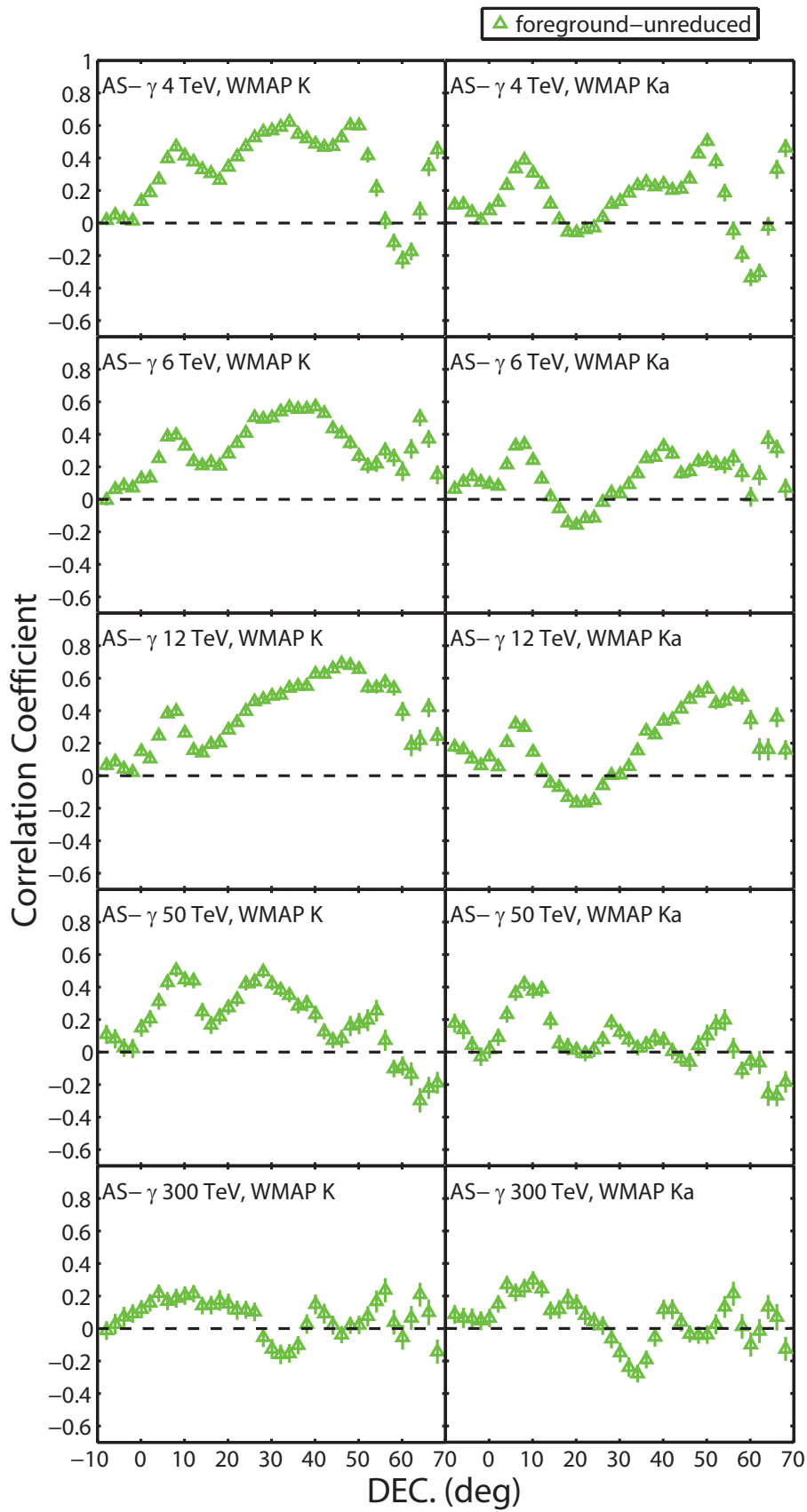}
\caption{Correlation coefficients (green triangles) of the AS-$\gamma$ intensity maps (the left five panels
in Figure \ref{Fig1}) with the WMAP nine-year foreground-unreduced data at
K and Ka bands (the right top two panels in Figure \ref{Fig1}) as functions
of declinations. \label{Fig3}}
\end{figure}

\begin{figure}[H]
\centering
\includegraphics[scale=0.8]{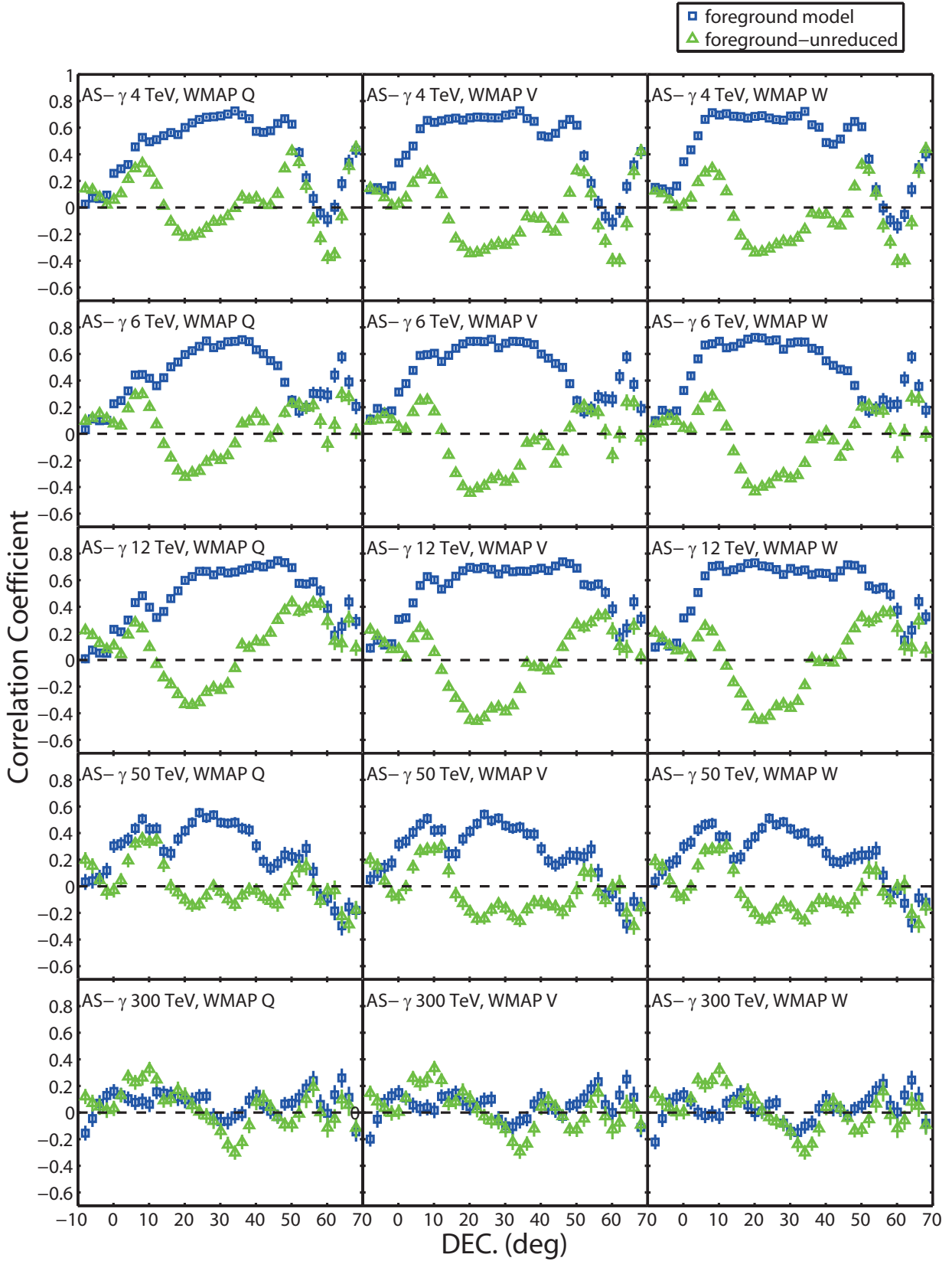}
\caption{Similar to Figure \ref{Fig3}, correlation coefficients (green triangles and blue squares) of the AS-$\gamma$ intensity maps (the left five panels in Figure \ref{Fig1}) with the WMAP nine-year foreground-unreduced data at Q ,V and W bands (the right bottom three panels in Figure \ref{Fig1}) and the WMAP template-based foreground models (the right three panels in Figure \ref{Fig2}). \label{Fig4}}
\end{figure}

With the pre-processing described in Section \ref{sec21}, we can obtain the CR maps in different energy bands and the WMAP maps in different frequency bands, as shown in Figures \ref{Fig1} and \ref{Fig2}. Then we use the CR maps to do correlation analysis with the WMAP foreground-unreduced maps and the total WMAP template-based foreground models with the PCC method described in Section \ref{sec22}, and obtain the curves of PCCs and their errors as functions of declinations as shown in Figures \ref{Fig3} and \ref{Fig4}.

The WMAP foreground-unreduced data contain the CMB and the total WMAP foregounds, and cover all five frequency bands (K, Ka, Q, V, and W). From the curves of PCCs between the CR and WMAP foreground-unreduced data shown in Figures \ref{Fig3} and \ref{Fig4}, we find that:
a) These PCCs are positive in K band and fall to negative as frequencies
increase till V and W bands. In the declination range of about $15^{\circ}$ to
$55^{\circ}$, where the CR maps only have the loss-cone structures, these
correlations fall off very fast as frequencies increase; whereas in
the declination range of about $-10^{\circ}$ to $15^{\circ}$, where
the CR maps have both the tail-in and loss-cone structures, they fall off more slowly as
frequencies increase;
b) These PCCs become weak for CR energy bands of 50 and 300 TeV.

The total WMAP template-based foreground models only cover three frequency bands (Q, V, and W). From the curves of PCCs between the CR data and total WMAP template-based models, we find that:
a) These PCCs are positive, and increase slightly with increasing WMAP frequencies till V and W bands;
b) These PCCs become weak for CR energy bands of 300 TeV.

\begin{figure}[H]
\centering
\includegraphics[scale=0.8]{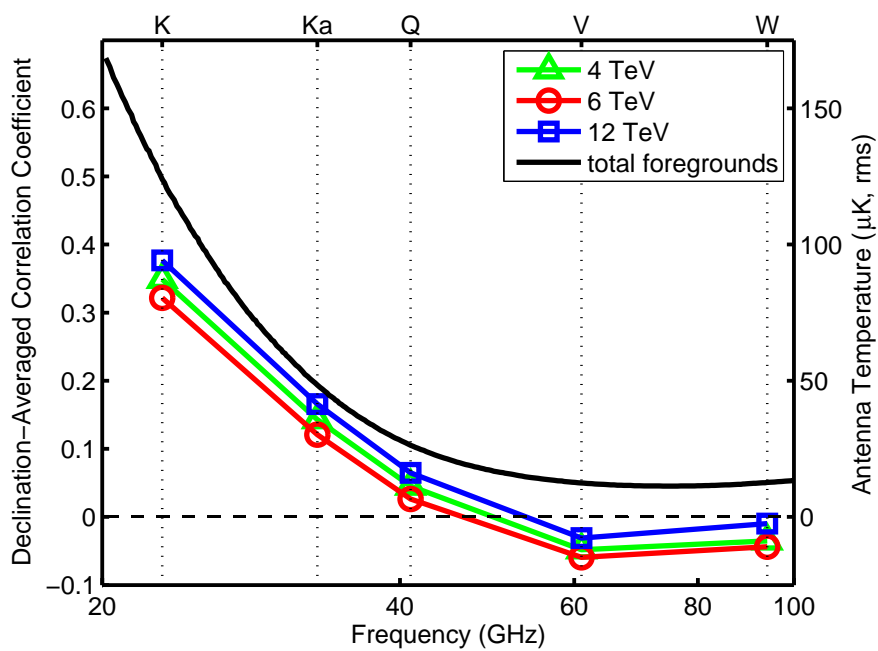}
\caption{Spectra of the declination-averaged correlation coefficients between the WMAP foreground-unreduced data and the CR maps at 4, 6 and 12 TeV (green triangles, red circles and blue squares), and spectra of total foreground models (black solid curves) outside of the KQ75y9 mask. \label{Fig5}}
\end{figure}

The anisotropic structures (loss-cone and tail-in) in the CR maps fade with increasing energies; this explains why PCCs become weak at higher energies. The spectra of the declination-averaged correlation coefficients for the different CR energy bands (4, 6 and 12 TeV) between the CR and WMAP foreground-unreduced data are similar to the spectra of total WMAP foregrounds (see Figure 22 in \citealt{ben13}) as shown in Figure \ref{Fig5}, so the CR maps may be used to trace the foregrounds in the WMAP data. Moreover, the significant positive PCCs between the CR data and total WMAP template-based foreground models show that they have correlations of spatial distribution; these correlations are more likely caused by the same physical mechanisms rather than accident, because the TeV CR can be significantly influenced by the Galactic environment (e.g., magnetic field) and the WMAP foregrounds are mainly Galactic components. Hence, at least some TeV CR we observe should originate from the Milky Way.

\subsection{Correlation betweeen CR and MEM Data\label{sec32}}

\begin{figure}[H]
\centering
\includegraphics[scale=0.8]{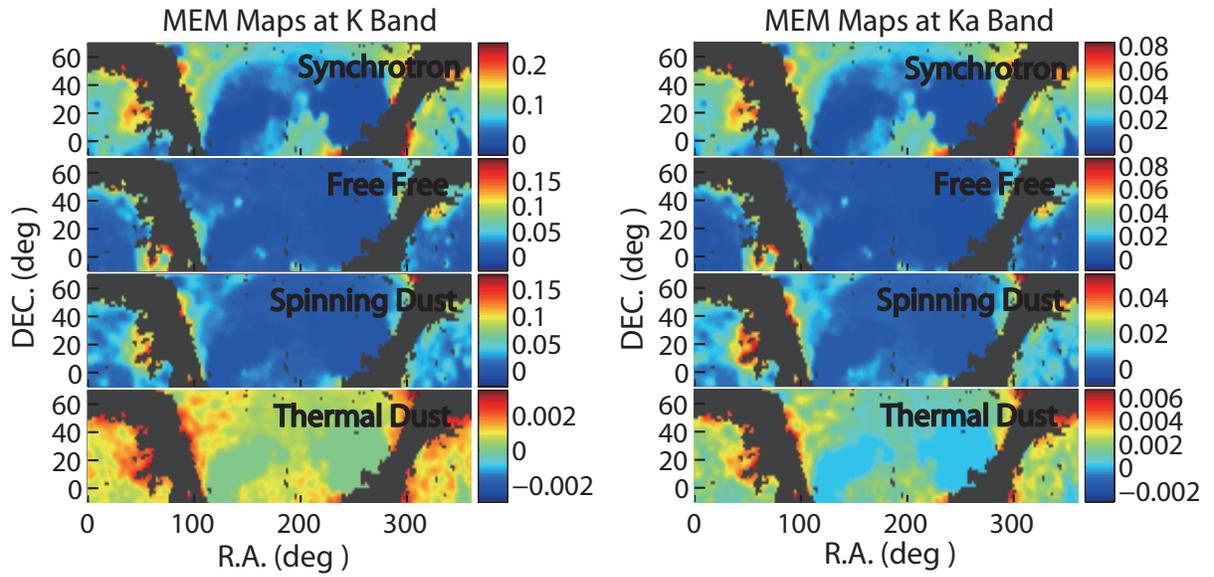}
\caption{Maximum Entropy Method (MEM) maps (synchrotron, free-free, spinning dust, and thermal dust foregrounds) at K and Ka bands. These maps are smoothed to an angular resolution of $5^{\circ}$. \label{Fig6}}
\end{figure}

\begin{figure}[H]
\centering
\includegraphics[scale=0.8]{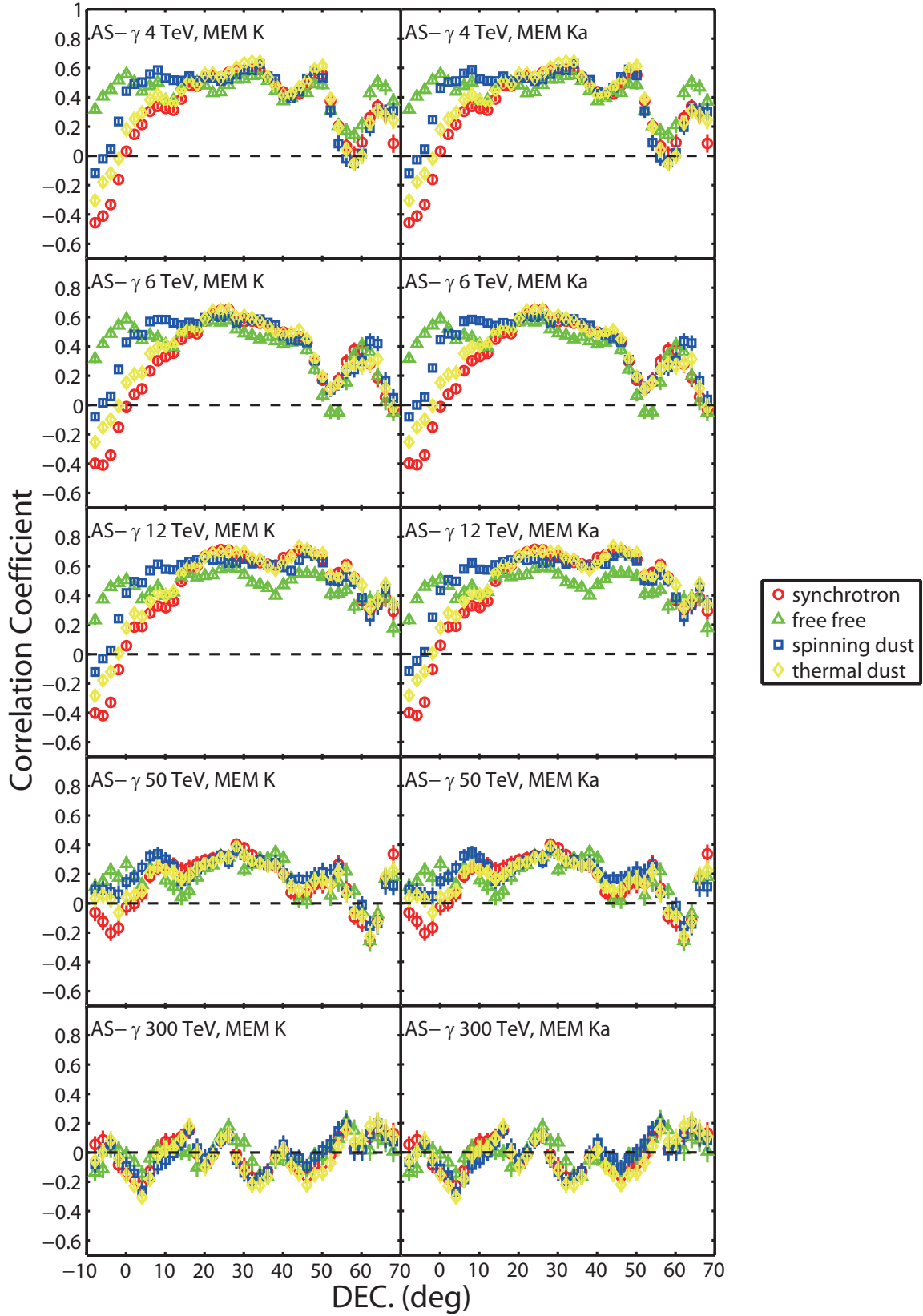}
\caption{Similar to Figure \ref{Fig3}, correlation coefficients (red circles, green triangles, blue squares and yellow diamonds) of the AS-$\gamma$ intensity maps (the left five panels in Figure \ref{Fig1}) with the synchrotron, free-free, spinning dust, and thermal dust MEM maps at K and Ka bands (the eight panels in Figure \ref{Fig6}). \label{Fig7}}
\end{figure}

The total WMAP template-based foreground models only cover Q, V and W bands; however, the Maximum Entropy Method (MEM) foreground models (synchrotron, free-free, spinning dust and thermal dust) enable us to do correlation analysis between the CR and these foregrounds to cover all WMAP frequency bands. The MEM generates foregrounds by using the smoothed WMAP maps minus the ILC map as input and some foreground maps as priors; the priors are the Haslam 408 MHz map \citep{has81,has82} for synchrotron, the H$\alpha$ map described in \cite{fin03} and \cite{ben13} for free-free, the SFD 100 micron map \citep{sch98} for spinning dust, and the FDS predicted 94 GHz thermal dust map \citep{fin99}. With the pre-processing described in Section \ref{sec21}, we can obtain the MEM foreground maps at K and Ka bands, as shown in Figure \ref{Fig6}. Then we use the CR maps to do correlation analysis with these MEM maps with the PCC method described in Section \ref{sec22}, and obtain the curves of PCCs and their errors as functions of declinations as shown in Figure \ref{Fig7}.

From the curves of PCCs between the CR and MEM maps, we find that:
a) These PCCs are almost all positive except three foregrounds (synchrotron, spinning dust and thermal dust) in the declination range of $-10^{\circ}$ to $0^{\circ}$;
b) These curves of PCCs at K band are identical with those at Ka band;
c) The PCC curves of four types of foregrounds have relatively significant differences in the declination range of $-10^{\circ}$ to $15^{\circ}$;
d) These PCCs become weak at 300 TeV.

Similar to that in Section \ref{sec31}, the PCCs become weak at high CR energy bands, due to the weak CR anisotropy at higher CR energies. In addition, the identical curves of PCCs at K and Ka bands suggest that each foreground's spatial distribution remain unchanged. In fact the WMAP team generated maps of a type of MEM foreground at each frequency band by multiplying this type of foreground's template by frequency dependent and pixel independent coefficient, i.e., there are the same spatial distribution for a type of MEM foreground map among different frequency bands; hence, the curves of PCCs at Q, V and W bands are identical with those at K or Ka band. Compared with the PCC curves of the declination range of $> 15^{\circ}$ (only cover the loss-cone structure), there are more significant differences in the declination range of $< 15^{\circ}$ (cover the loss-cone and tail-in structures) among four types of foregrounds; the loss-cone and tail-in features may have some physical differences, or the loss-cone features are different between these two declination ranges.

In Section \ref{sec31}, the PCC curves of the total WMAP template-based foreground models with the CR maps are slightly different among Q, V and W bands. Actually, each type of WMAP template-based foreground at all frequency bands also have identical spatial distribution; however, the proportion of different type of foreground in the total foreground model changes with increasing frequency, so the spatial distribution of the total foreground model changes as well. The PCCs of the foreground models (MEM and WMAP template-based) with the CR maps are almost all significantly positive; these correlations are more likely causal and can be used to understand the WMAP foregrounds because the TeV CR and the WMAP foregrounds are all related to Galactic ISM. Hence, CRs may be a good tracer of foregrounds at all WMAP frequencies. Combined with the PCC curves of the WMAP foreground-unreduced data with the CR maps in Section \ref{sec31}, we can conclude that the total foregrounds in the WMAP foreground-unreduced data dominate at K and Ka bands while the CMB dominate at Q, V and W bands; hence, the spectra of the declination-averaged correlation coefficients agree with the spectra of total foregrounds. Moreover, the results of this section also suggest that the values of PCCs between the CMB and CR maps may be negative.

\section{FOREGROUND RESIDUALS IN WMAP DATA\label{sec4}}

Section \ref{sec3} shows that the CR data are useful for probing the WMAP foregrounds, so we can do correlation analysis with the WMAP foreground-reduced data. However, possible accidental correlations between the CR and CMB sky maps have an effect on finding out the foreground residuals, due to cosmic variance. Hence, we need theoretical CMB sky maps to estimate possible impacts of accidental correlations, and the latest Planck sky map is also helpful.

\subsection{Cosmic Variance\label{sec41}}

The CMB sky map detected by the WMAP is one sample of the population including many possible CMB sky maps (e.g., observed from other galaxies) with identical cosmological parameters; therefore, the cosmic variance is used to measure the effect of the fluctuations of all possible CMB maps to the correlations between CR data and CMB sky maps. Hence, we need large amounts of mock CMB sky maps to reflect intrinsic fluctuation of CMB in spatial distribution and obtain the cosmic variance.

SYNFAST, i.e, a program in HEALPix software, is used to create 10,000 mock CMB sky maps (smoothed to $5^{\circ}$) with the same angular power spectra generated by CAMB tool; here we use the cosmological parameters provided by the WMAP nine-year cosmological results (see Table 17 in \citealt{ben13}). Every mock CMB sky map has a random spatial distribution of temperature, and then 10,000 correlation coefficients ($r_{xy}^{\rm \uppercase\expandafter{\romannumeral2}}$ calculated with Equation (\ref{rxy})) are calculated between the CR data and 10,000 mock CMB sky maps at each declination. Finally, $r_{xy}^{\rm \uppercase\expandafter{\romannumeral2}}$ can be transformed to $z_{xy}^{\rm \uppercase\expandafter{\romannumeral2}}$ with Equation (\ref{zxy}), and then 10,000 $z_{xy}^{\rm \uppercase\expandafter{\romannumeral2}}$ at each declination approximately follow a normal distribution with the mean $<z_{xy}^{\rm \uppercase\expandafter{\romannumeral2}}>$ and the standard deviation $\sigma_{z_{xy}}^{\rm \uppercase\expandafter{\romannumeral2}}$. Therefore, the mean $<r_{xy}^{\rm \uppercase\expandafter{\romannumeral2}}>$, the one-sigma $\sigma_{r_{xy}}^{\rm \uppercase\expandafter{\romannumeral2}}$ and the two-sigma $\sigma_{r_{xy},2 \sigma}^{\rm \uppercase\expandafter{\romannumeral2}}$ can be derived from $<z_{xy}^{\rm \uppercase\expandafter{\romannumeral2}}>$, $\sigma_{z_{xy}}^{\rm \uppercase\expandafter{\romannumeral2}}$, and 2 $\sigma_{z_{xy}}^{\rm \uppercase\expandafter{\romannumeral2}}$ by the inverse Fisher transformation, respectively; here the cosmic variance can be represented by $\sigma_{r_{xy}}^{\rm \uppercase\expandafter{\romannumeral2}}$, while $\sigma_{r_{xy}}^{\rm \uppercase\expandafter{\romannumeral1}}$ represent the measurement variance of an instrument (e.g., the detectors in the WMAP observatory).

\begin{figure}[H]
\centering
\includegraphics[scale=0.8]{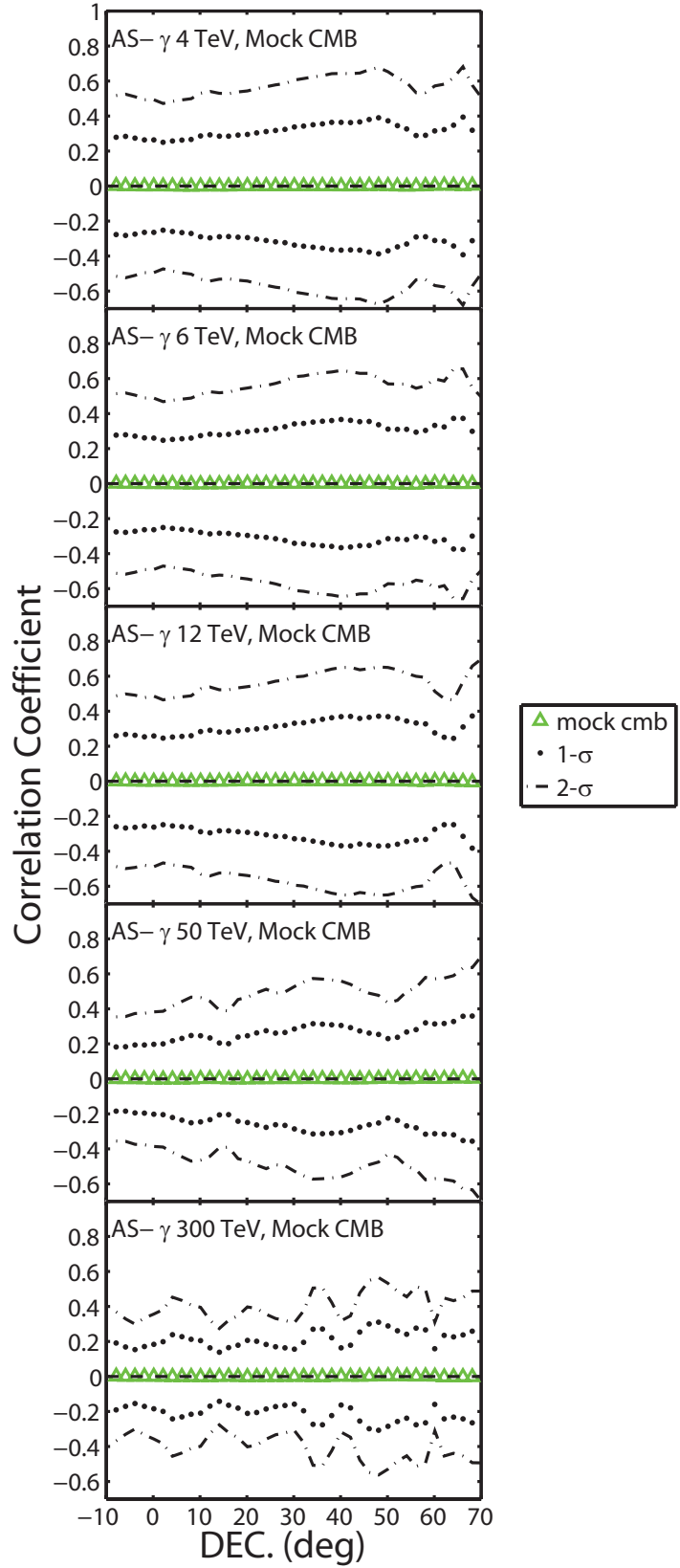}
\caption{The statistical results of 10,000 sets of correlation coefficients between the CR maps and the mock CMB maps: the mean values (green triangles), 1-$\sigma$ cosmic variance lines (dotted lines) and 2-$\sigma$ cosmic variance lines (dot-dashed lines). \label{Fig8}}
\end{figure}

If there are only accidental correlations, $<r_{xy}^{\rm \uppercase\expandafter{\romannumeral2}}>$ approaches to zero with increasing sample numbers; otherwise, the CR and CMB maps have causal correlations. In addition, if we want to ensure that the observed CMB sky map is credible, the measurement variance must be much smaller than the cosmic variance (i.e., $\sigma_{r_{xy}}^{\rm \uppercase\expandafter{\romannumeral2}}$ $\gg$ $\sigma_{r_{xy}}^{\rm \uppercase\expandafter{\romannumeral1}}$). Figure \ref{Fig8} shows the statistics results of 10,000 sets of PCCs between the CR and mock CMB maps: the mean $<r_{xy}^{\rm \uppercase\expandafter{\romannumeral2}}>$, 1-$\sigma$ ($\sigma_{r_{xy}}^{\rm \uppercase\expandafter{\romannumeral2}}$) lines and 2-$\sigma$ ($\sigma_{r_{xy},2 \sigma}^{\rm \uppercase\expandafter{\romannumeral2}}$) lines. We find that all mean values (i.e., $<r_{xy}^{\rm \uppercase\expandafter{\romannumeral2}}>$) are close to zero;
therefore, we conclude that the CR and CMB maps only have accidental correlations, as expected.

\subsection{Correlation betweeen CR and WMAP Foreground-reduced Data\label{sec42}}

\begin{figure}[H]
\centering
\includegraphics[scale=0.7]{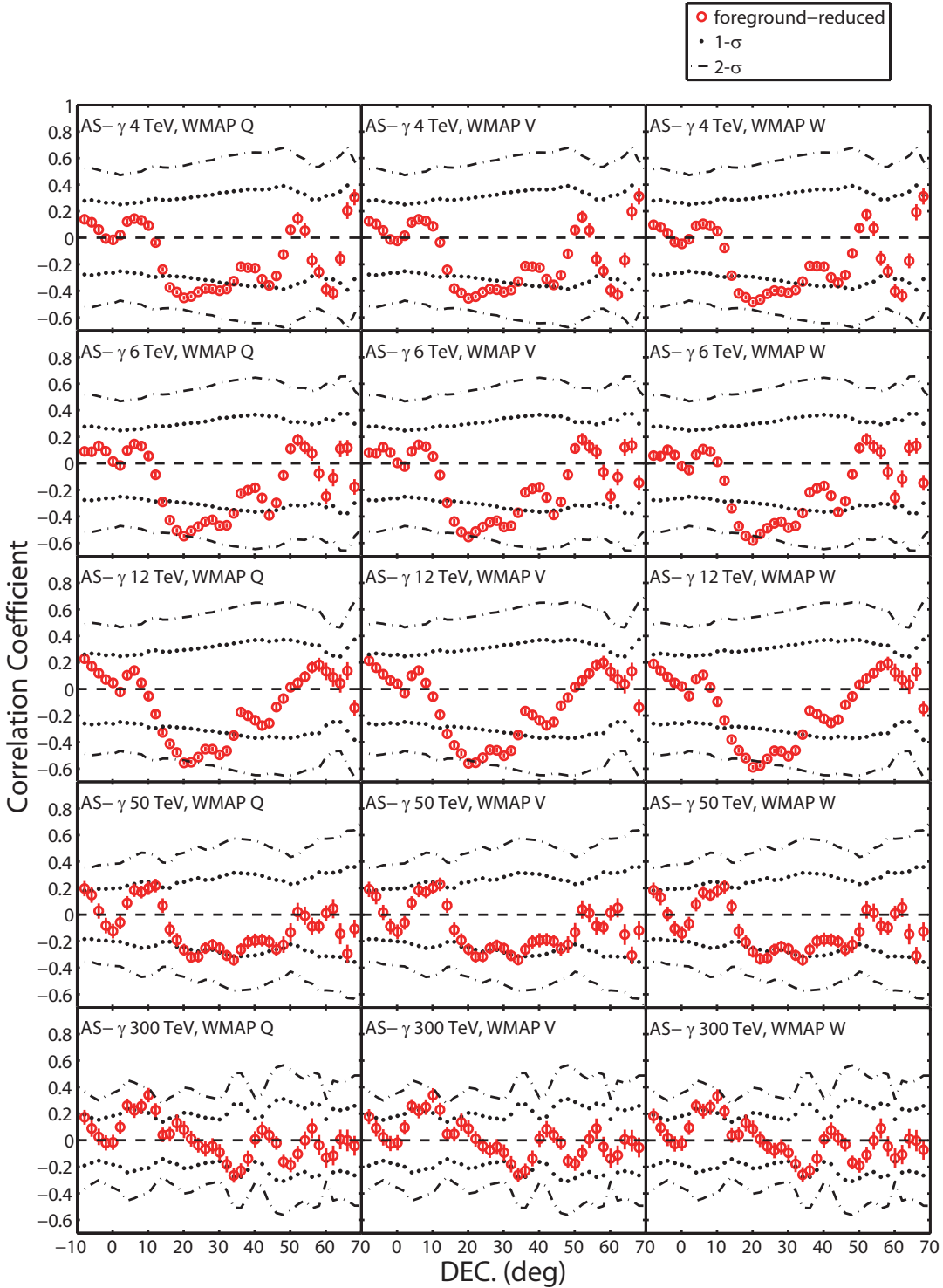}
\caption{Correlation coefficients (red circles) of the AS-$\gamma$ intensity maps (the left five panels
in Figure \ref{Fig1}) with the WMAP nine-year foreground-reduced data at
Q, V and W bands (the left three panels in Figure \ref{Fig2}) as functions
of declinations. We use 1-$\sigma$ cosmic variance lines (dotted lines) to divide non-significant correlations from significant correlations and 2-$\sigma$ cosmic variance lines (dot-dashed lines) to divide significant correlations from highly significant correlations. \label{Fig9}}
\end{figure}

The WMAP foreground-reduced data only cover Q, V and W bands, and are used to do correlation analysis with the CR maps with the PCC method described in Section \ref{sec22}. The curves of PCCs and their errors with declinations are shown in Figure \ref{Fig9}; here
1-$\sigma$ lines and 2-$\sigma$ lines are drawn in figures.

For the curves of PCCs between the CR and WMAP foreground-reduced data, we find that:
a) $\sigma_{r_{xy}}^{\rm \uppercase\expandafter{\romannumeral2}}$ $\gg$ $\sigma_{r_{xy}}^{\rm \uppercase\expandafter{\romannumeral1}}$;
b) These PCCs are negative and significant ($|r_{xy}^{\rm \uppercase\expandafter{\romannumeral1}}|$ $>$ $\sigma_{r_{xy}}^{\rm \uppercase\expandafter{\romannumeral2}}$) from 4 to 12 TeV in the
declination range of about $15^{\circ}$ to $35^{\circ}$, and become weak
for CR energy bands of 50 and 300 TeV;
c) These PCCs have almost no change from Q to W bands at the same CR energy bands.

Therefore, we can conclude that: (1) the measurement variance is much smaller than the cosmic variance, as described in Section \ref{sec41}; and (2) there are anticorrelations between the CR and WMAP foreground-reduced maps.

\subsection{Correlation betweeen CR and Planck Data\label{sec43}}

\begin{figure}[H]
\centering
\includegraphics[scale=0.8]{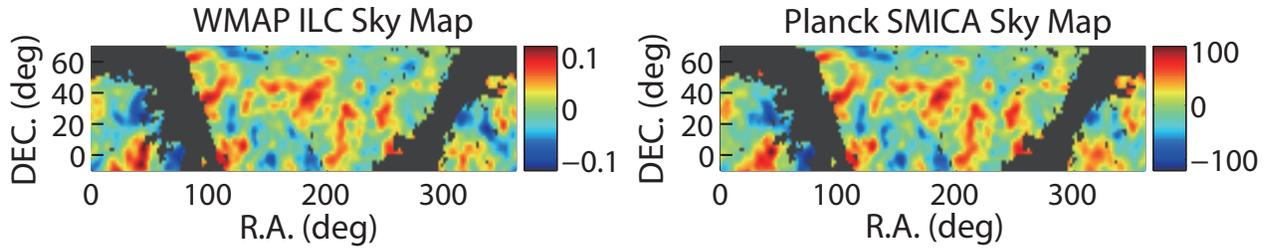}
\caption{The WMAP ILC sky map (reduced-galaxy map produced via linear combination technique) and the Planck SMICA sky map (reconstructed CMB map as a linear combination). These two maps are smoothed to an angular resolution of $5^{\circ}$. \label{Fig10}}
\end{figure}

\begin{figure}[H]
\centering
\includegraphics[scale=0.8]{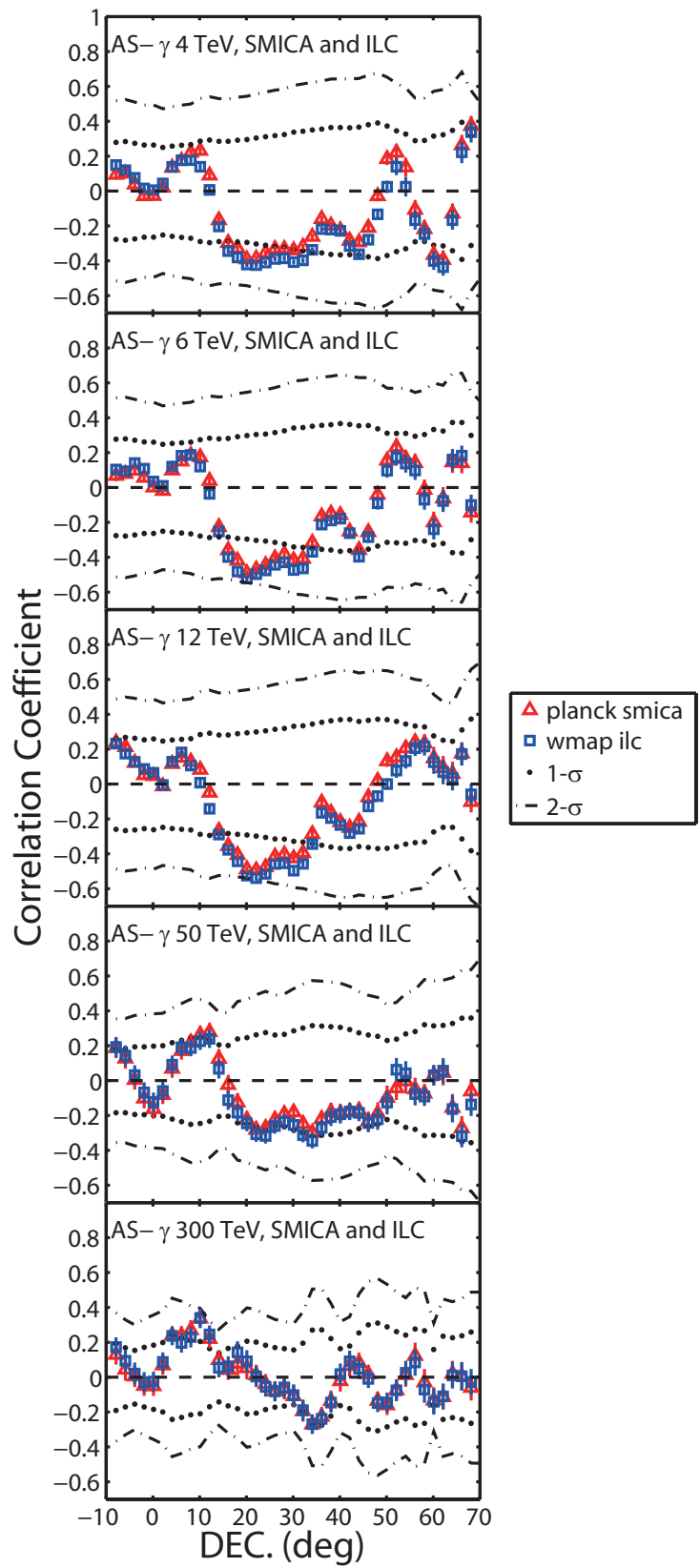}
\caption{Similar to Figure \ref{Fig9}, correlation coefficients (blue squares and red triangles) of the AS-$\gamma$ intensity maps (the left five panels in Figure \ref{Fig1}) with the WMAP ILC sky map (the left panel in Figure \ref{Fig10}) and the Planck SMICA sky map (the right panel in Figure \ref{Fig10}). \label{Fig11}}
\end{figure}

The Planck space mission has released higher SNR maps \citep{pla13,xii13}, so that we can compare the WMAP ILC sky map with the Planck SMICA sky map. These are CMB maps formed by linearly combining data from all WMAP bands or all Planck channels with weighting such that foreground emission is approximately canceled. The causal correlations with CR data caused by foreground residuals may result in different PCCs of these two maps. The WMAP ILC and Planck SMICA maps are processed with the pre-processing described in Section \ref{sec21} as shown in Figure \ref{Fig10}, and the curves of PCCs and their errors with declinations are shown in Figure \ref{Fig11}.

In Figure \ref{Fig11}, we find that:
a) $\sigma_{r_{xy}}^{\rm \uppercase\expandafter{\romannumeral2}}$ $\gg$ $\sigma_{r_{xy}}^{\rm \uppercase\expandafter{\romannumeral1}}$;
b) These PCCs are negative and significant ($|r_{xy}^{\rm \uppercase\expandafter{\romannumeral1}}|$ $>$ $\sigma_{r_{xy}}^{\rm \uppercase\expandafter{\romannumeral2}}$) from 4 to 12 TeV in the
declination range of about $15^{\circ}$ to $35^{\circ}$, and become weak
for CR energy bands of 50 and 300 TeV.
Therefore, we can conclude that: (1) the measurement variance is much smaller than the cosmic variance, as described in Section \ref{sec41}; and (2) there are anticorrelations between the CR and WMAP ILC (or Planck SMICA) maps.

The PCC curves of both WMAP ILC and Planck SMICA maps have no significant difference, and the PCCs with the CR data of the WMAP ILC map are stronger slightly than those of the Planck SMICA map in the declination range of about $15^{\circ}$ to $35^{\circ}$. Hence, we cannot conclude that the Planck SMICA map have more serious problems about foreground residuals than the WMAP ILC map.

\subsection{Probability of Accidental Correlation\label{sec44}}

The reduced foreground sky maps (WMAP foreground-reduced, WMAP ILC and Planck SMICA maps) and the CR maps have significant ($|r_{xy}^{\rm \uppercase\expandafter{\romannumeral1}}|$ $>$ $\sigma_{r_{xy}}^{\rm \uppercase\expandafter{\romannumeral2}}$) anticorrelations from 4 to 12 TeV in the declination range of about $15^{\circ}$ to $35^{\circ}$; here the loss-cone structures are located in this declination range. However, we cannot reject the hypothesis that these anticorrelations are accidental, before we make quantitative analysis to calculate probabilities that these anticorrelations are accidental.

There are 10,000 mock CMB sky maps in Section \ref{sec41}, and we can find out the number that the correlations of the mock CMB and CR maps are stronger than those of the reduced foreground and CR maps, i.e., the absolute value of the sum of weighted PCCs between the mock CMB and CR maps $|\sum{r_{xy}^{\rm \uppercase\expandafter{\romannumeral2}}}W_{xy}|$ and the absolute value of the sum of weighted PCCs between the reduced foreground and CR maps $|\sum{r_{xy}^{\rm \uppercase\expandafter{\romannumeral1}}}W_{xy}|$ satisfy: $|\sum{r_{xy}^{\rm \uppercase\expandafter{\romannumeral2}}}W_{xy}|$ $\geq$ $|\sum{r_{xy}^{\rm \uppercase\expandafter{\romannumeral1}}}W_{xy}|$; here the weight $W_{xy}$ ($\propto$ $\rm cos(Dec.)$) is used to account for the variation of the solid angle of the sampled sky with declinations. Then the probability of accidental correlation (AC) can be calculated by
\beq P_{\rm AC}=\frac{N_{\rm AC}}{10,000}, \label{num}
\eeq
where $N_{\rm AC}$ is the number of maps with $|\sum{r_{xy}^{\rm \uppercase\expandafter{\romannumeral2}}}W_{xy}|$ $\geq$ $|\sum{r_{xy}^{\rm \uppercase\expandafter{\romannumeral1}}}W_{xy}|$ among 10,000 mock CMB sky maps in Section \ref{sec41}.

We need to choose a reasonable declination range before calculating $P_{\rm AC}$. Full available declination range can be used to describe accidental correlation in the declination range of about $-10^{\circ}$ to $70^{\circ}$; however, narrower declination range is more reasonable when foreground residuals only dominate at local ranges. Therefore, the significant ($|r_{xy}^{\rm \uppercase\expandafter{\romannumeral1}}|$ $>$ $\sigma_{r_{xy}}^{\rm \uppercase\expandafter{\romannumeral2}}$) anticorrelations in the declination range of about $15^{\circ}$ to $35^{\circ}$ are worth special attention. Table \ref{Tab1} lists $P_{\rm AC}$ of WMAP (Q to W band and ILC) and Planck SMICA maps from 4 to 12 TeV in three different declination ranges: `A' for full available declination range, `B' for $-10^{\circ}$ to $35^{\circ}$ declination range (cover the loss-cone and tail-in structures) and `C' for $15^{\circ}$ to $35^{\circ}$ declination range (only cover the loss-cone structure); here the probabilities of 50 and 300 TeV CR energy bands are not calculated, because the CR maps at these energy bands are nearly isotropic. We can conclude that: (1) the foreground residuals exist in the WMAP foreground-reduced maps with $\sim$ 95\% probability at scales of $20^{\circ}$, but accidental correlations with $\sim$ 5\% probability cannot be excluded; (2) compared with the WMAP ILC map, the Planck SMICA map has lower foreground residual levels because of the larger $P_{\rm AC}$; (3) the data in the declination range of $< 15^{\circ}$ have great impact on the probabilities, because `B' range's maximum $P_{\rm AC}$ difference (23\%) is much larger than `A' range's (7\%) or `C' range's (10\%), i.e., `B' range's $P_{\rm AC}$ has the biggest change with the WMAP frequency or CR energy bands; (4) the values of $P_{\rm AC}$ decrease with increasing WMAP frequency or CR energy bands.

\section{SUMMARY AND DISCUSSION\label{sec5}}

We have first performed correlation analysis of the AS-$\gamma$ CR data with the foreground-contained sky maps (i.e., the WMAP foreground-unreduced maps, the WMAP template-based foreground models and the MEM foreground models) to confirm that the CR can be used to probe foregrounds in the WMAP data. Next, we generate 10,000 mock CMB sky maps to describe the cosmic variance, and then analyzed the significant anticorrelations of the CR data with the WMAP foreground-reduced maps. Moreover, the latest Planck SMICA map is used to compare with the WMAP ILC map. In order to understand whether the foreground residuals exist, we obtain the probabilities of accidental or causal correlations. The main conclusions of this paper are as following.

1) The spectra of the declination-averaged correlation coefficients for the different CR energy bands (4, 6 and 12 TeV) between the CR and WMAP foreground-unreduced data is similar to the spectra of total WMAP foregrounds. Hence, the CR data can be used to trace the WMAP foregrounds.

2) The correlations of the CR data with the WMAP template-based foreground models and the MEM foreground models (synchrotron, free-free, spinning dust and thermal dust) suggest the Galactic origin of at least some TeV CRs.

3) The correlations of the WMAP foregrounds with the anisotropic structures (loss-cone and tail-in structures) in the CR maps show that the existence of these structures may be related to some foreground emissions (e.g., synchrotron emissions).

4) The correlations of the CR data with four types of foregrounds have relatively significant differences in the declination range of $< 15^{\circ}$, and the maximum $P_{\rm AC}$ difference of $-10^{\circ}$ to $35^{\circ}$ declination range (23\%) is much larger than the maximum $P_{\rm AC}$ difference of $15^{\circ}$ to $35^{\circ}$ declination range (10\%); hence, the CR data in the declination range of $< 15^{\circ}$ have great impact on the correlation results.

5) The cosmic variances are far greater than the measurement variances, so the WMAP foreground-reduced maps are credible for making correlation analysis with the CR data.

6) The foreground residuals exist in the WMAP foreground-reduced maps with high probabilities $\sim$ 95\% at scales of $20^{\circ}$, but accidental correlations $\sim$ 5\% cannot be excluded.

7) The Planck SMICA map has lower foreground residual levels than the WMAP ILC map, because the accidental correlation probabilities of the former are greater than those of the latter.

Three main components, including CMB, CR and foregrounds, are studied in this work. Since the total WMAP template-based foreground models cannot contain all foreground emissions, so some foregrounds may still exist in the WMAP foreground-reduced map, i.e, the final WMAP CMB map may not be clean. Hence, the CR data are used to find out the foreground residuals, because the CR data we use have two advantages: (1) the TeV CR have close relationship with the foregrounds; and (2) the TeV CR have no any physical relationship with the CMB from the early universe at redshift z $\sim$ 1100. However, the accidental correlations due to cosmic variance between the CR and CMB maps (i.e., only the correlations of spatial distributions, not a physical relationship) have an effect on probing the foregrounds; nevertheless we can calculate the probabilities of the accidental (or causal) correlations.

The anisotropic structures (loss-cone and tail-in structures) play an important role in the correlation analysis. For both the total WMAP template-based foreground models and the MEM foreground models (synchrotron, free-free, spinning dust and thermal dust), the correlations with the CR maps are related to the loss-cone structure and the tail-in structure, i.e., the foreground maps have some similar structures. These structures may result from (1) the propagation of the cosmic rays in some regions of the Milky Way changes the spatial distribution of foreground emissions; and (2) the spatial distribution of the CRs and the foregrounds are all affected by some common physical mechanisms. However, cosmic rays are not directly involved in free-free, spinning dust and thermal dust emission mechanisms, and even the synchrotron emission associated with CR anisotropy would be negligible compared to the synchrotron emission from the general ISM; here CR anisotropy is most likely the local features (near the solar system), so the associated synchrotron emission originate over a short path length rather than entire path length through the Galaxy. One possible explanation is that: there exist many CR producing anisotropic structures throughout the Milky Way, those distant structures produce a roughly isotropic CR map while some local structures lead to anisotropy in the CR map. Then these local structures, which contain gas and dust, also generate all kinds of microwave foregrounds correlated with CR anisotropy. Moreover, each anisotropic structure has its own characteristics, including motions and components (gas, dust, magnetic filed, supernova remnant, etc.), and then different types of structures may have different effect (inhibition or promotion) on CR propagation. Hence, the distinction between the tail-in and loss-cone structures may be due to different types of anisotropic structures.

Whether the foreground residuals exist in the WMAP foreground-reduced map has not been solved completely, because we cannot exclude the accidental correlations with high confidence. Hence, further works are needed: a) better foreground templates and higher SNR CMB sky map are helpful; b) the formation mechanism of the anisotropic structures in the CR map is still an open question, and further studies about the loss-cone and tail-in structures may be useful to understand our correlation results; c) more available absolute calibration of CR detectors and more CR data over larger sky areas can help us to better probe the foreground residuals.

\begin{table}
\begin{center}
\caption{Probabilities of accidental correlations between the WMAP Foreground-reduced Maps and the CR Maps\label{Tab1}}
\begin{tabular}{|c|c|c|c|c|c|}
\hline

              & Q band  &  V band &  W band &  ILC & SMICA \\
              & ($\%$)  & ($\%$) & ($\%$) & ($\%$) & ($\%$) \\
\hline
\multicolumn{6}{|c|}{A: full available declination range} \\
\hline
4 TeV       & $37.7_{-7.3}^{+8.6}$   & $36.9_{-7.5}^{+8.2}$   & $32.2_{-7.1}^{+8.2}$  &  $41.8_{-8.2}^{+8.8}$  &  $55.3_{-9.6}^{+10.1}$ \\
6 TeV       &  $35.5_{-8.3}^{+9.6}$       &  $35.1_{-8.3}^{+9.6}$       &  $30.7_{-7.7}^{+8.8}$  &  $41.8_{-9.2}^{+10.6}$  &  $49.0_{-10.2}^{+11.0}$ \\
12 TeV      & $36.8_{-8.4}^{+10.1}$   & $36.1_{-8.2}^{+10.1}$  & $32.2_{-8.1}^{+9.1}$  & $43.5_{-9.8}^{+10.6}$ & $56.1_{-10.8}^{+12.0}$  \\
\hline
\multicolumn{6}{|c|}{B: $-10^{\circ}$ to $35^{\circ}$ declination range} \\
\hline
4 TeV       & $44.7_{-5.6 }^{+5.9}$   & $43.0_{-5.6}^{+5.7}$   & $36.0_{-5.4}^{+5.3}$  &  $50.4_{-6.0}^{+6.7}$  &  $59.4_{-6.8}^{+7.1}$ \\
6 TeV       &  $33.6_{-5.6}^{+6.6}$       &  $31.8_{-5.3}^{+6.5}$       &  $26.0_{-4.4}^{+5.4}$  &  $40.2_{-6.6}^{+7.1}$  &  $44.6_{-7.0}^{+7.2}$ \\
12 TeV      & $29.1_{-5.2}^{+6.0}$   & $27.6_{-5.0}^{+5.7}$  & $22.4_{-4.4}^{+5.1}$  & $35.1_{-6.0}^{+6.9}$ & $43.8_{-7.3}^{+7.6}$  \\
\hline
\multicolumn{6}{|c|}{C: $15^{\circ}$ to $35^{\circ}$ declination range} \\
\hline
4 TeV       & $14.8_{-1.7}^{+2.1}$   & $14.1_{-1.6}^{+2.1}$   & $12.1_{-1.6}^{+1.5}$  &  $16.2_{-2.1}^{+2.0}$  &  $23.7_{-2.5}^{+2.6}$ \\
6 TeV       & $8.0_{-1.3}^{+1.5}$      &  $7.6_{-1.3}^{+1.5}$       &  $6.3_{-1.2}^{+1.3}$  &  $9.1_{-1.5}^{+1.6}$  &  $12.9_{-1.8}^{+2.5}$ \\
12 TeV      & $6.2_{-1.2}^{+1.4}$   & $5.8_{-1.2}^{+1.3}$  & $4.6_{-0.9}^{+1.3}$  & $7.3_{-1.3}^{+1.9}$ & $11.7_{-1.8}^{+2.5}$  \\

\hline
\end{tabular}
\end{center}
\end{table}

\acknowledgments
The anonymous referee is thanked for his/her very insightful and constructive comments and suggestions, which allowed us to improve the manuscript significantly. We are very grateful to the Tibet AS-$\gamma$ collaboration for providing their CR maps in electronic format and
especially Dr. Yi Zhang for helping us in understanding the data. We acknowledge the use of LAMBDA, support for which
is provided by the NASA/GSFC, the software packages HEALPix (http://healpix.sourceforge.net/), developed by A.J. Banday, M. Barthelmann, K.M. Gorski, F.K. Hansen, E. Hivon, and B.D. Wandelt, and CAMB (http://camb.info), developed by A. Challinor and A. Lewis. SNZ acknowledges partial funding support by Directional Research Project of the Chinese Academy of Sciences under project No. KJCX2-YW-T03 and by the
National Natural Science Foundation of China under grant Nos. 10821061, 10733010,10725313, and by 973 Program of
China under grant 2009CB824800.

\begin{thebibliography}{}

\bibitem[Abbasi et al.(2011)]{abb11} Abbasi, R., et al. 2011, \apj, 740, 16

\bibitem[Abdo et al.(2009)]{abd09} Abdo, A.A., et al. 2009, \apj, 698, 2121

\bibitem[Aglietta et al.(2009)]{agl09} Aglietta, M., \& EAS-TOP Collaboration 2009, \apj, 692, L130

\bibitem[Amenomori et al.(2006)]{ame06} Amenomori, M., et al. 2006, Science, 314, 439

\bibitem[Bennett et al.(2003)]{ben03} Bennett, C.L., et al. 2003, \apjs, 148, 97

\bibitem[Bennett et al.(2013)]{ben13} Bennett, C.L., et al. 2013, \apjs, 208, 20

\bibitem[de Oliveira-Costa et al.(1999)]{deo99} de Oliveira-Costa, A.,
Tegmark, M., Guti$\acute{e}$rrez, C.M., Jones, A.W., Davies, R.D.,
Lasenby, A.N., Rebolo, R., \& Watson, R.A. 1999, \apj, 527, L9

\bibitem[de Oliveira-Costa et al.(2004)]{deo04} de Oliveira-Costa, A.,
Tegmark, M., Davies, R.D., Guti$\acute{e}$rrez, C.M., lasenby, A.N.,
Rebolo, R., \& Watson, R.A. 2004, \apj, 606, L89

\bibitem[Drury \& Aharonian(2008)]{dru08} Drury, L.O'C., \&  Aharonian, F.A. 2008, Astropart. Phys., 29, 420

\bibitem[Ferrie\`re(2001)]{fer01} Ferrie\`re, K.M. 2001, Reviews of Modern Physics, 73, 1031

\bibitem[Finkbeiner(1999)]{fin99} Finkbeiner, D.P., Davis, M., \& Schlegel, D.J. 1999, \apj, 524, 867

\bibitem[Finkbeiner(2003)]{fin03} Finkbeiner, D.P., 2003, \apjs, 146, 407

\bibitem[Finkbeiner(2004)]{fin04} Finkbeiner, D.P., 2004, \apj, 614, 186

\bibitem[Giacinti \& Sigl(2012)]{gia12} Giacinti, G., \& Sigl, G. 2012, Phys. Rev. Lett., 109, 1101

\bibitem[Ginzburg \& Syrovatskii(1965)]{gin65} Ginzburg, V.L., \& Syrovatskii, S.I. 1965, Annu. Rev. Astron. Astrophys., 3, 297

\bibitem[Guillian et al.(2007)]{gui07} Guillian, G., et al. 2007, Phys. Rev. D, 75, 062003

\bibitem[Haslam et al.(1981)]{has81} Haslam, C.G.T., et al. 1981, \aap, 100, 209

\bibitem[Haslam et al.(1982)]{has82} Haslam, C.G.T., et al. 1982, \aaps, 47, 1

\bibitem[Hinshaw et al.(2007)]{hin07} Hinshaw, G.F., et al. 2007, \apjs, 170, 288

\bibitem[Hinshaw et al.(2013)]{hin13} Hinshaw, G.F., et al. 2013, \apjs, 208, 19

\bibitem[Hirata et al.(2008)]{hir08} Hirata, C.M., et al. 2008, Physical Review D., 78, 3520

\bibitem[Liu \& Zhang(2006)]{liu06} Liu, X., \& Zhang, S.N. 2006, \apj, 636, L1

\bibitem[Nagashima \& Mori(1976)]{nag76} Nagashima, K., \& Mori, S. 1976,
in Proc. Int. Cosmic Ray Symp. on High Energy Cosmic Ray Modulation,
Univ. of Tokyo, Tokyo, Japan, 326

\bibitem[Nagashima et al.(1998)]{nag98} Nagashima, K., Fujimoto, K., \& Jacklyn, R.M. 1998,
J. Geophys. Res. 103, 17429

\bibitem[Padmanabhan et al.(2005)]{pad05} Padmanabhan, N., et al. 2005, Physical Review D., 72, 3525

\bibitem[Planck Collaboration et al.(2013a)]{pla13} Planck Collaboration, et al. 2013a, \aap, 571, A1

\bibitem[Planck Collaboration et al.(2013b)]{xii13} Planck Collaboration, et al. 2013b, \aap, 571, A12

\bibitem[Schlegel et al.(1998)]{sch98} Schlegel, D.J., Finkbeiner, D.P., \& Davis, M. 1998, \apj, 500, 525

\bibitem[Verschuur(2007)]{ver07} Verschuur, G.L. 2007, \apj, 671, 447

\bibitem[Wibig \& Wolfendale(2005)]{wib05} Wibig, T., \& Wolfendale, A.W. 2005, \mnras, 360, 236

\bibitem[Zhang et al.(2010)]{zha10} Zhang, J., Zhang, Y., Cui, S., \& Argo-YbJ Collaboration 2010, 38th COSPAR Scientific Assembly, 38, 2707

\end{thebibliography}
\end{document}